\documentclass[usenatbib]{mn2e}
\usepackage{epsfig}
\usepackage{amsmath,amssymb}
\usepackage{aas_macros}

\usepackage[dvips]{color}
\definecolor{rev}{rgb}{0.8,0.0,0.0}
\definecolor{cut}{rgb}{0.5,0.5,0.5}

\newcommand{\Kel}{$\rm SMSS \ J0313-6708$}

\newcommand{\Caf}{$\rm SDSS \ J1029+1729$}

\newcommand{\percc}{{\rm cm^{-3}}}
\newcommand{\um}{{\rm \mu m}}
\newcommand{\E}[1]{\times 10^{#1}}

\newcommand{\Enstatite}{{\rm MgSiO_3}}
\newcommand{\Forsterite}{{\rm Mg_2SiO_4}}

\newcommand{\nH}{n_{{\rm H}}}
\newcommand{\mH}{m_{{\rm H}}}

\newcommand{\Esn}{E_{{\rm SN}}}

\newcommand{\XH}{X_{{\rm H}}}

\newcommand{\Msun}{{\rm M}_{\bigodot}}

\newcommand{\Mni}{M(^{56}{\rm Ni})}

\newcommand{\Mpr}{M_{{\rm pr}}}

\newcommand{\erg}{\rm erg}

\defcitealias{Chiaki15}{C15}
\defcitealias{Marassi14}{M14}
\defcitealias{Nozawa13}{NK13}

\newcommand{\ACcr}{A_{\rm cr}({\rm C})}
\newcommand{\abA}[1]{A({\rm {#1}})}
\newcommand{\abH}[1]{{\rm [{#1}/H]}}
\newcommand{\aby}[1]{y({\rm {#1}})}
\newcommand{\abFe}[1]{{\rm [{#1}/Fe]}}

\newcommand{\rfC}{r_{\rm C}^{\rm cool} / f_{\rm C,C} }
\newcommand{\rfMg}{r_{\rm Sil}^{\rm cool} / f_{\rm Sil,Mg} }

\title[Classification of EMP stars]
      {Classification of extremely metal-poor stars: absent region in $\abA{C}$-$\abH{Fe}$ plane and the role of dust cooling}

\author[G. Chiaki et al.]
{Gen Chiaki,$^{1}$\thanks{E-mail: chiaki@center.konan-u.ac.jp}
Nozomu Tominaga$^{1}$ and
Takaya Nozawa$^{2}$
\\
$^{1}$Department of Physics, Konan University,
8-9-1 Okamoto, Kobe, 658-0072, Japan \\
$^{2}$Division of Theoretical Astronomy, National Astronomical Observatory 
of Japan, Mitaka, Tokyo 181-8588, Japan
}

\begin{document}

\date{}

\pagerange{\pageref{firstpage}--\pageref{lastpage}} \pubyear{2015}

\maketitle

\label{firstpage}

\begin{abstract}
Extremely metal-poor (EMP) stars are the living fossils with records of chemical enrichment history at the early epoch of galaxy formation.
By the recent large observation campaigns, statistical samples of EMP stars have been obtained.
This motivates us to reconsider their classification and formation conditions.
From the observed lower-limits of carbon and iron abundances of $\ACcr \sim 6$ and $\abH{Fe}_{\rm cr} \sim -5$
for C-enhanced EMP (CE-EMP) and C-normal EMP (CN-EMP) stars, we confirm 
that gas cooling by dust thermal emission is indispensable for the fragmentation of their parent clouds to form such low-mass, i.e., 
long-lived stars, and that the dominant grain species are carbon and silicate, respectively.
We constrain the grain radius $r_i^{\rm cool}$ of a species $i$ and condensation efficiency $f_{ij}$ of a key element $j$
as $\rfC = 10 \ \um$ and $\rfMg = 0.1 \ \um$ to reproduce $\ACcr$ and $\abH{Fe}_{\rm cr}$, which
give a universal condition $10^{\abH{C}-2.30}+10^{\abH{Fe}} > 10^{-5.07}$ for the formation of every EMP star.
Instead of the conventional boundary $\abFe{C} = 0.7$ between CE- and CN-EMP stars,
this condition suggests a physically meaningful boundary $\abFe{C}_{\rm b} = 2.30$ 
above and below which carbon and silicate grains are dominant coolants, respectively.
\end{abstract}



\begin{figure*}
\includegraphics[width=8.5cm]{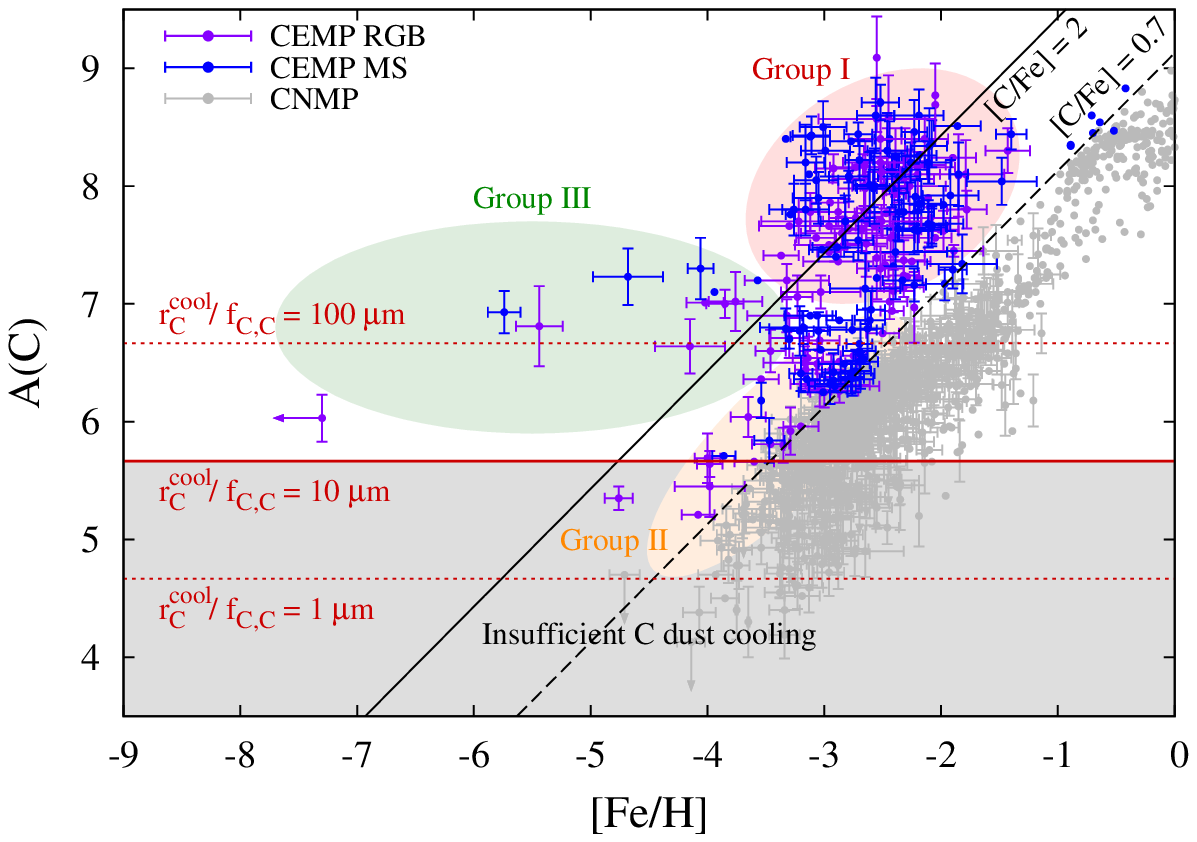}
\includegraphics[width=8.5cm]{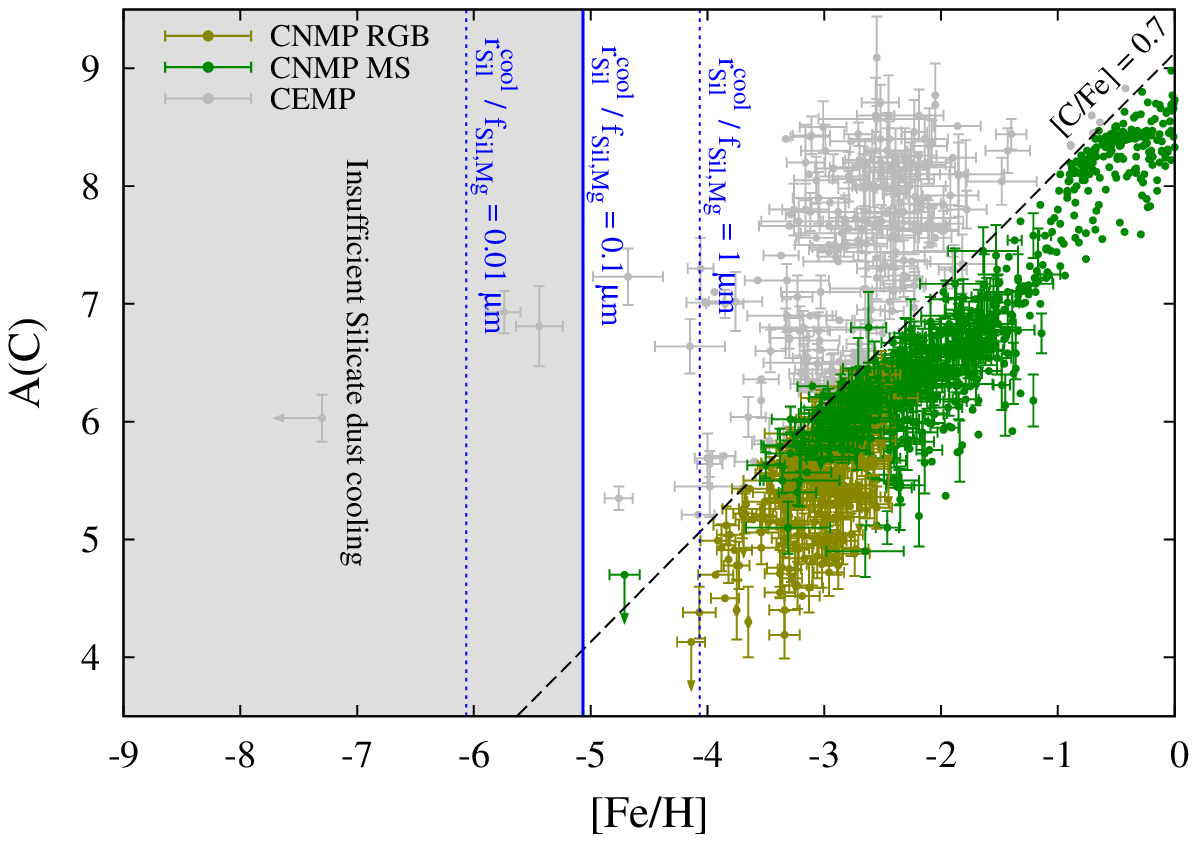}
\caption{
Carbon abundance $\abA{C}$ as a function of metallicity $\abH{Fe}$ of 
observed CEMP stars (left panel) and CNMP stars (right panel) retrieved from SAGA database
\citep[][http://sagadatabase.jp/]{Suda08, Suda11, Suda17, Yamada13}. 
The dashed line represents $\abFe{C} = 0.7$ as the boundary between CEMP and CNMP stars 
\citep{Beers05, Aoki07}.
CEMP stars are further divided into three groups indicated by the ellipses in the left panel \citep{Yoon16}, and 
the solid line ($\abFe{C} = 2.0$) indicates the apparent boundary between Groups II and III.
We show the critical carbon (Equation (\ref{eq:C_crit})) and iron (Equation (\ref{eq:Fe_crit}))
abundances by red and blue lines, respectively, for various $\rfC$ and $\rfMg$.
}
\label{fig:Fe_C_EMP}
\end{figure*}


\section{INTRODUCTION}

Long-lived stars with metallicities lower than our neighborhood, so-called metal-poor (MP) stars, are
discovered in our Galaxy and nearby dwarf galaxies.
They are intensively studied as the clues to know the chemical evolution during the structure formation.
This approach is called Galactic archeology or near-field cosmology.
MP stars are classified into carbon-enhanced MP (CEMP)
and carbon-normal MP (CNMP) stars,
divided at the boundary conventionally defined as $\abFe{C} = 0.7$ 
\citep{Beers05, Aoki07}.\footnote{The ratio between abundances of elements $j$ and $k$ are expressed as
$[j/k] = \log [y(j)/y(k)] - \log [y_{\odot } (j) / y_{\odot } (k)]$ 
and $A(j) = \log \epsilon (j) = 12 + \log y(j)$ for an element $j$,
where $y(j)$ denotes the number fraction of $j$ relative to hydrogen nuclei.} 
CEMP stars are further divided into CEMP-no without the enhancement of
neutron-capture elements, and CEMP-r or -s with r- or s-process element
enhancement, respectively \citep{Beers05}.

By recent large observational campaigns,\footnote{e.g., 
the HK \citep{Beers85, Beers92},
Hamburg/ESO \citep{Christlieb03},
SEGUE \citep{Yanny09}, and
LAMOST survey \citep{Cui12, Deng12}.}
we can access the statistical samples of MP stars.
\citet{Yoon16} report that CEMP stars are apparently subdivided into
three groups on the
$\abA{C}$-$\abH{Fe}$ plane (Figure \ref{fig:Fe_C_EMP}).
While CEMP Group I stars residing in $-3.7 < \abH{Fe} < -1.2$ and $7.0 < \abA{C} < 9.0$ are
dominantly CEMP-s stars,
Group II (with $-5.0 < \abH{Fe} < -2.5$ and $5.0 < \abA{C} < 7.0$) and 
Group III (with $\abH{Fe} < -3.5$ and $6.0<\abA{C} < 7.5$) stars are 
mainly CEMP-no stars.
Since almost all Group I stars show binary feature, they are considered to have acquired the gas 
rich with C and $n$-capture elements from their evolved companions
\citep{Suda04}.
On the other hand, the physical explanation of distinction between Group II and III stars has not been made so far.

MP stars with $\abH{Fe} < -3$ including CEMP Group II and III stars
are particularly called extremely metal-poor (EMP) stars.
The lower-limits of their elemental abundances indicate the existence of the critical metallicity above which their
parent clouds become unstable to fragment into small gas clumps through efficient gas cooling by heavy elements
so that low-mass stars which survive until the present day are likely to be formed
\citep{Bromm03, Frebel05}.
Recent theoretical studies have shown that cooling by dust thermal emission
is crucial to form low-mass fragments with $\sim 0.1 \ \Msun$ \citep{Omukai00, Schneider03}.

\citet{Marassi14} and \citet{Chiaki15} predict that dust grains
are important commonly for the formation of C-enhanced EMP (CE-EMP) and C-normal EMP (CN-EMP) stars.
They show that the dominant grain species are carbon and silicate, respectively, 
and estimate the critical C and Si abundances.
However, they resort to theoretical models of grain formation 
because the properties of grains such as radius and condensation efficiency
can not be directly measured.
Further, their analyses are based on the conventional classification of EMP stars, and do not
explain the difference between Group II and III stars.

In this Letter, we reconsider the classification and formation conditions of 
EMP stars.
From Figure~\ref{fig:Fe_C_EMP}, 
we point out three interesting features; 
(1) no EMP stars have so far been observed in the region
of $\abA{C} < 6$ and $\abH{Fe} < -5$,
(2) Group III and II stars are distributed in the regions with
high ($\abFe{C} > 2$) and moderate ($\abFe{C} < 2$) C-enhancement, respectively,
and 
(3) the distribution of Group II stars appears continuously connected
with CN-EMP stars.

We derive grain properties of carbon and silicate from the feature (1), and
then present a formation condition applicable to every EMP star.
Comparing the contributions of carbon and silicate grains to gas cooling,
we propose the physically motivated boundary between EMP stars 
whose formation could be derived mainly by carbon and silicate grains.
This boundary can simultaneously explain the features (2) and (3).
Throughout this Letter, we use the solar abundance
of \citet{Asplund09} as
$A_{\odot}({\rm C}) = 8.43$,
$A_{\odot}({\rm Mg}) = 7.60$, and 
$A_{\odot}({\rm Fe}) = 7.50$.

\begin{figure*}
\includegraphics[width=12cm]{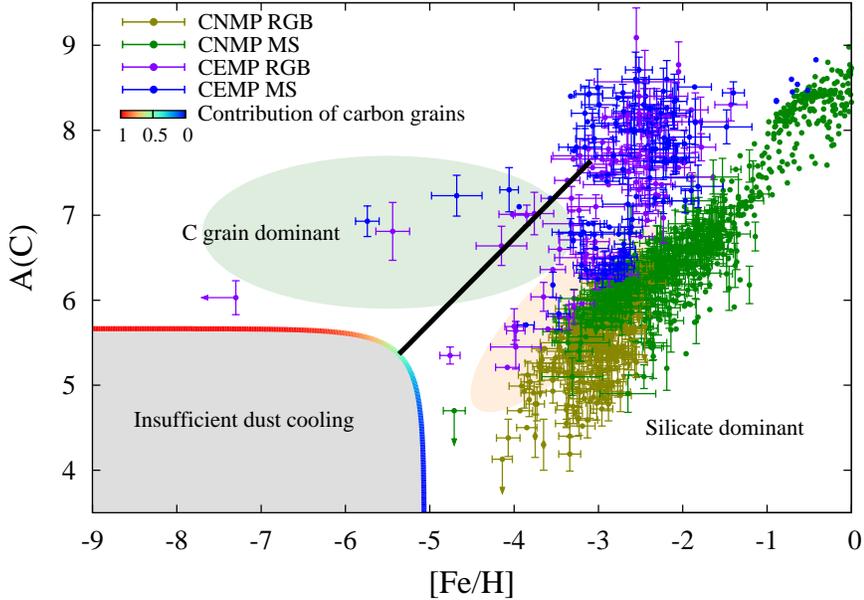}
\caption{
Carbon abundance $\abA{C}$ as a function of $\abH{Fe}$ of 
Galactic halo MP stars ever observed.
The formation of these long-lived, low-mass stars would be restricted
by C and Fe abundances above which gas cooling by carbon and silicate grains
exceeds gas compressional heating (Equation (\ref{eq:C_Sil_crit})) 
indicated by the curve colored from blue to red with increasing fraction of the carbon grain cooling efficiency.
The black line shows the boundary $\abFe{C} _ {\rm b} = 2.30$ (Equation (\ref{eq:C_Sil_trans}))
above and below which carbon and silicate grains are dominant coolants for the formation of EMP stars, respectively.
}
\label{fig:Fe_C_All}
\end{figure*}

\section{Critical elemental abundances}

The critical condition for cloud fragmentation can be described by the comparison of
gas cooling owing to dust thermal emission with
gas compressional heating \citep{Schneider12Crit}.
With condensation efficiency $f_{ij}$ of a key element $j$ onto a grain species $i$ 
and a characteristic grain radius $r_i^{\rm cool}$,
the fragmentation condition can be written using the number abundance $y(j)$ as
\begin{equation}
\sum _i \frac{3 \mu _{ij} f_{ij} \XH}{4 \varsigma _i r_i^{\rm cool}} y(j) > 1.4\E{-3} \ {\rm cm^{2} \ g^{-1}},
\label{eq:dust_crit}
\end{equation}
where $\mu _{ij}$ denotes the molecular weight of a monomer, and
$\varsigma _i$ is the bulk density of a grain \citep{Chiaki15}.\footnote{
We here consider spherical grains.
The radius $r_i^{\rm cool}$ is defined as $\langle r^3 \rangle _i / \langle r^2 \rangle _i $
characterizing the efficiency of gas cooling, where
$\langle x \rangle _i = \int x \varphi _i (r) dr$ is the average of 
a physical quantity $x$ weighted by the size distribution $\varphi _i(r)$ of a grain species $i$.
Equation (\ref{eq:dust_crit}) is given at the gas density $\nH = 10^{14} \ \percc$ and
temperature $T=1000$ K where dust cooling is dominant over gas compressional heating
in clouds with $\abH{Fe} \sim -5$ \citep{Schneider12Crit}.
$\XH$ is the mass fraction of hydrogen nuclei, and $\XH =0.75$ throughout this Letter.}
This indicates that, once the key element and its abundance $y(j)$ are specified,
we can put a constraint on $r_i^{\rm cool}$ and $f_{ij}$ in a form of $r_i^{\rm cool} / f_{ij}$,
which we hereafter call the effective grain radius.

\subsection{Critical C abundance and property of carbon grains}

We first derive the effective radius $\rfC$ for carbon grains.
For carbon grains, the key element is carbon, and
$\mu _{\rm C, C} = 12$ and $\varsigma _{\rm C} = 2.23 {\rm g \ cm^{-3}}$.
Then, from Equation (\ref{eq:dust_crit}), we can obtain the critical carbon abundance above which gas cooling by carbon grains
exceeds gas compressional heating as
\begin{equation}
\ACcr = 5.67 + \log \left( \frac{\rfC }{ 10 \ \um }  \right).
\label{eq:C_crit}
\end{equation}
The horizontal red lines in the left panel of Figure \ref{fig:Fe_C_EMP} show the critical carbon abundances
for $\rfC = 1$, $10$, and $100 \ \um$ from bottom to top.
CEMP Group III stars 
are distributed in a range of $\abA{C} > 6.0$ over a wide range of $\abH{Fe}$ ($-7 \lesssim \abH{Fe} \lesssim -4$).
Hence, $\ACcr = 6$ can be taken as the lower-limit of C abundances for CEMP Group III stars.
Then, the effective grain radius is estimated to be $\rfC = 21.6 \ \um$.
Taking the statistical uncertainties into consideration, we here adopt $\rfC = 10 \ \um$.
Below the horizontal line corresponding to $\rfC = 10 \ \um$ (shaded region in Figure \ref{fig:Fe_C_EMP}), 
carbon dust cooling is inefficient to induce cloud fragmentation,
and long-lived low-mass stars are unlikely to form. 
Some of CEMP Group II stars have lower carbon abundances than $\ACcr = 6.0$,
which is discussed in the subsequent sections.

\subsection{Critical Fe abundance and property of silicate grains}

Next, we constrain the property of the other major grain species, silicate.
For silicates, we consider enstatite ($\Enstatite$) 
with its key element being Mg, and
$\mu _{\rm \Enstatite, Mg} = 100$ and $\varsigma _{\Enstatite} = 3.21 {\rm g \ cm^{-3}}$.
The result is unchanged for forsterite ($\Forsterite$)
whose key element is Si.
From Equation (\ref{eq:dust_crit}), we can calculate the critical condition where
gas cooling by silicate grains overcomes the compressional gas heating as
\begin{equation}
\abH{Mg} _{\rm cr} = -4.70 + \log \left( \frac{\rfMg }{ 0.1 \ \um }  \right).
\end{equation}
We convert this critical Mg abundance to the critical Fe abundance, using the average
abundance ratio $\abFe{Mg} = 0.368$ for stars with $\abH{Fe} < -1$
(from SAGA database), as
\begin{equation}
\abH{Fe} _{\rm cr} = -5.07 + \log \left( \frac{\rfMg }{ 0.1 \ \um }  \right).
\label{eq:Fe_crit}
\end{equation}
The vertical blue lines in the right panel of Figure \ref{fig:Fe_C_EMP} indicate
the critical Fe abundances for $\rfMg = 0.01$, $0.1$, and $1 \ \um$ from left to right.
To realize the critical abundance $\abH{Fe} _{\rm cr} = -5.0$ suggested
by the distribution of CN-EMP stars, the effective grain radius $\rfMg = 0.081 \ \um$ is required.
We set the fiducial value of the effective grain radius as $\rfMg = 0.1 \ \um$,
which is smaller than that of carbon grains by two orders of magnitude.

\subsection{Combined criterion for CE- and CN-EMP star formation}

In the previous sections, we have considered separately the contributions of carbon
and silicate grains to gas cooling, and derived their effective grain radii.
In reality, both the grain species can contribute simultaneously to gas cooling
in collapsing clouds.
In this case, Equation (\ref{eq:dust_crit}) is reduced to
\[
\frac{3.03}{\rfC} \aby{C} + \frac{17.5}{\rfMg} \aby{Mg} > 1.4\E{-3}.
\]
Using $\rfC = 10 \ \um$ and $\rfMg = 0.1 \ \um$ derived in the previous sections gives
\begin{equation}
10^{\abH{C} - 2.30} + 10^{\abH{Fe}} > 10^{-5.07}.
\label{eq:C_Sil_crit}
\end{equation}
The critical condition is shown by the colored curve in Figure \ref{fig:Fe_C_All}.
The shaded region below the curve can successfully explain the region where no stars 
have been observed so far.
Also, the figure suggests that the distribution of CEMP Group II stars shows the lower-limit of Fe abundance
at $\abH{Fe} = -5$ as that of CN-EMP stars.

\subsection{Boundary between CE- and CN-EMP stars}
\label{sec:C_Fe_tr}

Equating the first and second terms in the left hand side of Equation (\ref{eq:C_Sil_crit}),
we can define the transition line 
on which the contributions of carbon and silicate grains to gas cooling are equal
as
\begin{equation}
\abFe{C} _{\rm b} = 2.30.
\label{eq:C_Sil_trans}
\end{equation} 
This condition is indicated by the black line in Figure \ref{fig:Fe_C_All}
above which gas cooling by carbon grain is dominant over that by silicates.
The boundary gives the physical explanation of the distinction between CEMP Group III
and Group II stars.

\begin{figure}
\includegraphics[width=8.5cm]{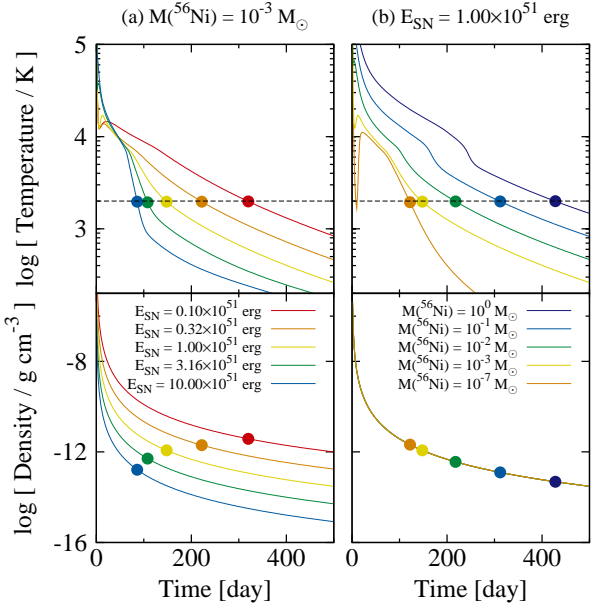}
\caption{
Time evolutions of gas temperature and 
density
(a) with a fixed $^{56}$Ni mass of $\Mni = 10^{-3} \ \Msun$ and (b) with a fixed explosion energy of $\Esn = 1\E{51} \ \erg$
at the mass coordinate where the largest carbon grains are formed.
The circles are plotted at the onset time of carbon grain formation when $T$ becomes below the condensation temperature
(2000 K) indicated by the black dashed line.
}
\label{fig:tnT}
\end{figure}

\begin{figure}
\includegraphics[width=8.5cm]{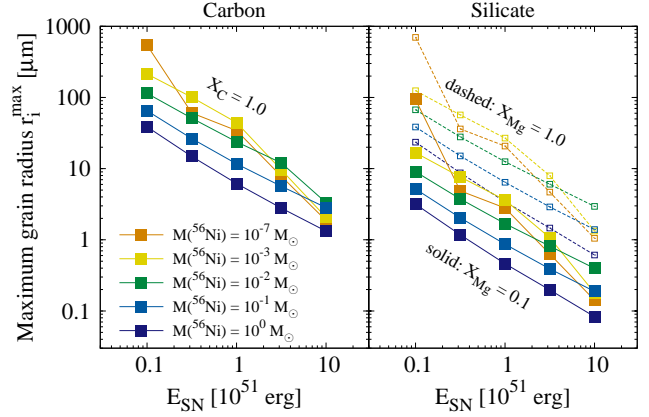}
\caption{
Maximum grain radius $r_i^{\rm max}$ of carbon (left) and silicate (right) grains
as a function of explosion energy $\Esn$ with various $^{56}$Ni masses.
For silicate grains, dashed and solid curves show the results with
mass fraction $X_{\rm Mg}$ of the key element Mg of 1.0 and 0.1, respectively.
}
\label{fig:Er}
\end{figure}

\section{Model calculations of grain properties in Pop III SN ejecta}

We have shown that the effective grain radius of carbon must
be larger than that of silicate by two orders of magnitude to reproduce the observed
distributions of EMP stars.
We in this section 
show this difference in grain radius 
with a dust formation model.
In the early Universe, dust grains are mainly supplied by supernovae (SNe) arising from
first-generation metal-free (Pop III) stars \citep{Todini01, Nozawa03}.
While the elemental abundances of CE-EMP stars are well
reproduced by faint core-collapse SNe (FSNe) with C
enhancement due to large fallback of Fe peak elements,
those of CN-EMP stars are reproduced by energetic core-collapse 
SNe (CCSNe) or hypernovae (HNe)
\citep{UmedaNomoto03, Limongi12, Marassi14, Tominaga14, Ishigaki14}.

To estimate the properties of newly formed dust,
we follow the 
temporal evolution of temperature $T$ and density $\rho$
of expanding SN ejecta with radiative transfer calculations including energy deposition
from radioactive decay of $^{56}$Ni \citep{Iwamoto00}.
Applying the approximation formulae of \citet{Nozawa13},
the average grain radius $r_i^{\rm ave} (M_{\rm R})$ and condensation efficiency $f_{ij}^{\rm ave} (M_{\rm R})$ 
are calculated at each mass coordinate $M_{\rm R}$
from the concentration $c_j^{\rm on} = X_j \rho / m_j$ of a key element $j$
and the cooling timescale $\tau _i^{\rm cool}$ at the onset time $t_i^{\rm on}$ of dust formation when
$T$ declines down to the condensation temperature
(2000 K and 1500 K for carbon and silicate grains, respectively).
A dominant fraction of grains will be formed
at the mass coordinate $M_{\rm R}^{\rm max}$ where $r_i^{\rm ave}$ becomes largest
because $c_i^{\rm on}$ marks maximum there.
We thus take $r_i^{\rm max} = r_i^{\rm ave} (M_{\rm R}^{\rm max})$ as the fiducial value of the grain radius.
At $M_{\rm R}^{\rm max}$, $f_{ij}^{\rm ave}$ turns to be $\sim 1$.
We take a progenitor model of the SN which reproduces the chemical composition 
of $\rm HE0557-4840$ with an intermediate C-enhancement $\abFe{C} = 1.68$ (Ishigaki14).
The mass-cut is $M_{\rm cut} = 5.65 \ \Msun$ and progenitor mass is $\Mpr = 25 \ \Msun$.
The masses of C and Mg atoms are $m_{\rm C} = 12 \mH$ and $m_{\rm Mg} = 24 \mH$,
where $\mH$ is the mass of a hydrogen nucleus.

We first see the results for carbon grains.
Since carbon grains are formed mainly in the layer rich with C,
we set $X_{\rm C} = 1.0$.
Figure \ref{fig:tnT} shows the temporal evolutions of gas temperature and density.
With a fixed $^{56}$Ni mass of $\Mni = 1\E{-3} \ \Msun$, 
temperature and density for higher explosion energy $\Esn$ decline more rapidly
due to the higher expansion rate, and
the time $t_{\rm C}^{\rm on}$ (indicated by circles) becomes earlier.
With a fixed $\Esn = 1\E{51} \ \erg$, 
temperature keeps higher, and $t_{\rm C}^{\rm on}$ becomes later for larger $\Mni$.
Figure \ref{fig:Er} shows the grain radius $r_{i}^{\rm max}$ as a function
of $\Esn$ with various $\Mni$.
For each $\Mni$, the grain radius is smaller for higher $\Esn$
because the gas density is smaller at $t_{\rm C}^{\rm on}$ by more rapid expansion.
For $\Mni < 0.01 \ \Msun$, $r_i^{\rm max}$ declines rapidly with $\Esn \gtrsim 1.0 \ \erg$
because $t_i^{\rm on}$ is coincident with the finishing time of the plateau phase when
the ejecta becomes optically thin and the temperature decline suddenly.
With shorter cooling timescale, the larger number of grain seeds form, i.e, the radius
of each grain becomes smaller.

With the same mass fraction $X_{\rm C} = X_{\rm Mg} = 1.0$, $r_i^{\rm max}$ for carbon and silicate grains
are similar for each $\Esn$ and $\Mni$.
However, in the formation region of silicate grains, Mg is dominated by O, and we set the fiducial
value as $X_{\rm Mg} = 0.1$,
following nucleosynthesis calculations \citep{Tominaga14}.
With $X_{\rm Mg} = 0.1$, the $r_{\rm Sil}^{\rm max}$ decreases by an order of magnitude
for each $\Esn$ and $\Mni$.

To reproduce the elemental abundance of the most iron-poor CE-EMP stars such as \Kel \ \citep{Keller14},
$\Esn \sim 10^{51} \ \erg$ and $\Mni \lesssim 10^{-3} \ \Msun$ are favored \citep{Marassi14, Ishigaki14}.
In our calculations, we estimate $r_{\rm C}^{\rm max} = 34.3 \ \um$ with $\Esn = 1.0\E{51} \ \erg$ and
$\Mni = 1\E{-7} \ \Msun$.
The abundance ratio of the most metal-poor star \Caf \ \citep{Caffau11} is reproduced by HN models
with $\Esn \sim 10^{52} \ \erg$ and $\Mni \sim 0.1 \ \Msun$ \citep{Tominaga14}.
We predict $r_{\rm Sil}^{\rm max} = 0.188 \ \um$ with $\Esn = 1.0\E{52} \ \erg$, $\Mni = 0.1 \ \Msun$, and $X_{\rm Mg} = 0.1$.
These values are consistent with $\rfC = 10 \ \um$ and $\rfMg = 0.1 \ \um$.

\section{Discussion}
\label{sec:discussion}

The observed lower-limits of C and Fe abundances of C-enhanced and C-normal EMP stars indicate that
these stars form through the fragmentation of their parent clouds by gas cooling 
owing to thermal emission of two major grain species, carbon and silicate, respectively.
We first derive the grain radius and condensation efficiency as $\rfC = 10 \ \um$ and
$\rfMg = 0.1 \ \um$ from the lower-limits of C and Fe abundances, respectively.
The tendency that carbon grains are larger than silicates is qualitatively explained
by our simple analyses of dust formation.
Carbon grains grow more efficiently than silicate
because the gas density remains higher at the time when temperature declines
to the condensation temperature with the smaller $\Esn$ and $\Mni$ favored for CE-EMP stars, and
because the mass fraction of C is higher than that of Mg in the dust formation region.

We can derive the critical condition for EMP star formation as Equation (\ref{eq:C_Sil_crit}),
which can well reproduce the region where no stars have so far been observed as indicated by
the shaded region in Figure \ref{fig:Fe_C_All}.
Then, we find that the dominant coolant switches from carbon to silicate
from above to below the boundary $\abFe{C}_{\rm b} = 2.30$,
which gives the physically motivated classification of CE- and CN-EMP stars.
This simultaneously explain the discrimination of CEMP Group II and Group III stars \citep{Yoon16}.
Opposite to Group III stars, the dominant coolant for the formation of CEMP Group II stars 
is silicate grains as CN-EMP stars.
Interestingly, the distribution of Group II stars is continuous with that of CN-EMP stars
as indicated by Figure \ref{fig:Fe_C_All}.



Our estimation of the effective grain radii is based on the observed elemental 
abundances of EMP stars.
Our model presented in Section 3 predicts the larger values $\rfC = 34.3 \ \um$ and $\rfMg = 0.188 \ \um$
for the fiducial cases.
Although we here take the maximum grain radius at the corresponding mass coordinate $M_{\rm R}^{\rm max}$, 
the mass fraction of smaller grains formed in other mass coordinates can non-negligible, and 
the average radius of grains will be smaller.
\citet{Marassi15} predict the smaller grain radii of $\rfC \lesssim 0.1 \ \um$ by their grain formation
models in FSN ejecta.
It is still possible that stars with lower elemental abundances, which is permitted by
smaller effective grain radii, are discovered by future observations.
In the current state, although the number of samples in the Galactic halo is large ($\sim 10^6$ stars),
statistics of EMP stars with $\abH{Fe} < -3$ is still small \citep{Hartwig15}.
As the number of EMP stars increases by future observations, 
the accuracy of the estimation of grain property and the boundary between CE- and 
CN-EMP stars presented in this Letter will get improved.

\section*{acknowledgments}

GC is supported by Research Fellowships of the Japan Society for
the Promotion of Science (JSPS) for Young Scientists.
This work is supported in part by the Grants-in-Aid 
for Young Scientists (S: 23224004) and 
for General Scientists (A: 	16H02168, C: 26400223) by JSPS.


\end{document}